\documentclass[aps,prd,reprint,nofootinbib,superscriptaddress,raggedbottom,floatfix]{revtex4-2}
\usepackage{amsmath,graphicx}
\usepackage[hidelinks]{hyperref}
\hypersetup{pdftitle={Origin of curvature and radius scaling in charged black-hole interiors},pdfauthor={Fu-Guo Yang}}
\begin{document}
\title{Curvature scaling and linear area response in charged black-hole interiors}
\author{Fu-Guo Yang}
\email{yang-fuguo@outlook.com}
\affiliation{International Centre for Theoretical Physics Asia-Pacific, University of Chinese Academy of Sciences, 100190 Beijing, China}

\begin{abstract}
We identify the mechanisms and range of validity of two scaling laws reported in charged black-hole interiors. In spherical evolutions with a neutral massless scalar, mass inflation produces a large internal mass $m$, followed by a plateau. Small scalar and electric corrections give $K\simeq48m^2/r^6$, where $r$ is the areal radius and $K$ the Kretschmann scalar. The local exponent is near six when the mass and relative curvature correction also vary slowly with radius. The exact bound $K\geq28q^4/(5r^8)$, with $q$ the charge, excludes a finite-coefficient $r^{-6}$ asymptote, while a large plateau mass delays electric corrections to $r\sim q^2/m$. For sufficiently large plateau mass, this classical regime can reach Planckian curvature before electric corrections become significant. At fixed coordinates and initial gauge, a linear response of $\rho=r^2$ to the initial scalar amplitude $A$ gives the half-power radius law near a simple zero. Where both mechanisms apply, $K/m^2\propto(A_*-A)^{-3}$, with $A_*$ the area zero. Full evolutions confirm the electric correction, support the half-power law at several observation times, and test the joint inverse-cube law.
\end{abstract}
\maketitle
\allowdisplaybreaks[2]

\section{Introduction}
Shao, Guo, Tian, and Zhang reported two scaling laws for charged black-hole interiors evolved from time-symmetric initial data \cite{Shao}. Along the reflection line $x=0$, the Kretschmann scalar $K$ was fitted by $K\propto r^{-6.02\pm0.02}$ in the areal radius $r$. When the initial scalar amplitude was varied, the radius at a fixed coordinate location approached zero approximately as the square root of the amplitude difference from a threshold. These results raise two questions: what controls the measured powers, and over what range does the near-six curvature law remain valid?

The internal mass is central to the curvature law. Scalar backreaction can drive mass inflation, identified near the Cauchy horizon by Poisson and Israel \cite{PoissonIsrael} and described by Ori's exact model \cite{Ori}. Nonlinear studies resolve Cauchy-horizon focusing and spacelike singularities \cite{BradySmith,BurkoCH,BurkoCore,GuoJoshi}, while sufficiently small spherical perturbations of Reissner--Nordstr\"om (RN) data with two asymptotically flat ends can evolve without a spacelike singularity \cite{Dafermos}. The half-power radius law presents a separate mechanism question. In Choptuik's critical collapse \cite{Choptuik}, the mass-scaling exponent is set by a growing mode of the critical solution \cite{Gundlach1997,Gundlach}. Reference~\cite{Shao} leaves open whether its radius law has a similar origin or follows from a simpler response to the initial amplitude.

We explain the curvature power through the mass plateau that follows rapid scalar-driven mass growth. A large mass suppresses electric corrections, and slow variation of both the mass and the relative curvature correction permits an exponent near six. The exact curvature bound excludes a finite-coefficient sixth-power asymptote at fixed nonzero charge. The electric terms nevertheless remain small until $r\sim q^2/m_0$, where $m_0$ is the plateau mass. Comparing this transition with the Planck curvature scale shows when the classical near-six regime can extend to very high curvature before electric corrections become important.

The radius law concerns how $\rho=r^2$ at a fixed coordinate point changes between solutions with different initial scalar amplitudes. At fixed charge, scalar profile, and initial coordinate gauge, a finite, nonzero derivative of $\rho$ with respect to the initial amplitude $A$ gives the half-power radius law near a simple zero. We test this response at a fixed coordinate location and repeat the comparison at several observation times.

The relation $K/m^2\simeq48/\rho^3$ connects the two mechanisms. If $\rho$ is also linear in $A_*-A$, where $A_*$ is its zero, the mass-normalized curvature obeys $K/m^2\propto(A_*-A)^{-3}$. We test this prediction by evaluating area, mass, and curvature at the same coordinate location in each solution. The area response fixes the coefficient of this prediction independently of the curvature data.

\section{Model and definitions}
\label{sec:model}
We consider four-dimensional Einstein gravity coupled to an electric Maxwell field and a real, neutral, massless scalar $\varphi$. With signature $(-,+,+,+)$ and units $G=c=1$, the action is
\begin{equation}
 I=\int d^4x\sqrt{-g}\left[\frac{R-F_{ab}F^{ab}}{16\pi}-\frac12\nabla_a\varphi\nabla^a\varphi\right],
 \label{eq:action}
\end{equation}
where $R$ is the Ricci scalar and $F_{ab}$ is the electromagnetic field strength. We use the spherical metric
\begin{equation}
 ds^2=E(-dt^2+dx^2)+r^2d\Omega^2,\quad E=e^{-2\sigma}>0.
 \label{eq:metric}
\end{equation}
Here $r$ is the areal radius, so a symmetry sphere has area $4\pi r^2$, and $d\Omega^2$ is the metric of the unit sphere. The common factor $E$ is allowed by the local conformal flatness of the two-dimensional time--radial metric. Radial null curves satisfy $dt=\pm dx$.

The Maxwell equation fixes a conserved charge $q$. In spherical symmetry, the Misner--Sharp mass equals the Hawking mass and is given by $M_{\rm MS}=r[1-\nabla_a r\nabla^a r]/2$ \cite{MisnerSharp,Hawking}. We use the charged mass convention $m=M_{\rm MS}+q^2/(2r)$. Substituting the metric~\eqref{eq:metric} gives
\begin{equation}
 \frac{r_x^2-r_t^2}{E}=1-\frac{2m}{r}+\frac{q^2}{r^2}.
 \label{eq:mass}
\end{equation}
Subscripts denote partial derivatives. In RN, $m$ equals the constant mass parameter $M$. With scalar matter it evolves and need not equal the exterior mass. We measure curvature by the Kretschmann scalar $K=R_{abcd}R^{abcd}$.

Introducing the squared areal radius $\rho=r^2$, we write the Einstein and scalar field equations as
\begin{align}
 \rho_{tt}-\rho_{xx}&=-2E\left(1-\frac{q^2}{\rho}\right),\label{eq:rho}\\
 \begin{split}
 \sigma_{tt}-\sigma_{xx}&=-\frac{\rho_t^2-\rho_x^2}{4\rho^2}-\frac{E}{\rho}\\
 &\quad+\frac{2Eq^2}{\rho^2}+4\pi(\varphi_t^2-\varphi_x^2),
 \end{split}\label{eq:sigma}\\
 \varphi_{tt}-\varphi_{xx}&=-\frac{\rho_t\varphi_t-\rho_x\varphi_x}{\rho}.
 \label{eq:scalar}
\end{align}
The remaining Einstein equations impose two constraints. The numerical evolutions retain all source terms in Eqs.~\eqref{eq:rho}--\eqref{eq:scalar}.

We measure lengths and masses in units of the initial RN mass parameter, setting $M=1$. Curvatures are correspondingly measured in units of $M^{-4}$. The initial data are time symmetric, with all initial time derivatives zero. Following Ref.~\cite{Shao}, we prescribe $E(0,x)$ from RN with $0<q<1$ and use the same function of $x$ for every amplitude at fixed charge. For each scalar amplitude the initial radius satisfies the radial constraint with
\begin{equation}
 r(0,0)=r_+=1+\sqrt{1-q^2},\quad r_x(0,0)=0.
\end{equation}
The initial scalar is $\varphi(0,x)=Af(x)$, where $A$ is its amplitude and
\begin{equation}
 f(x)=\tanh x\quad\hbox{or}\quad
 f(x)=\operatorname{erf}\!\left(\frac{\sqrt\pi x}{2}\right).
 \label{eq:profiles}
\end{equation}
Here $\operatorname{erf}$ is the error function. Both profiles are odd, have $f'(0)=1$, and approach $\pm1$. Each evolution uses one profile. At fixed $A$, this normalization changes the profile shape while keeping the gradient at $x=0$ and the asymptotic scalar values fixed. The geometry remains even in $x$. Thus $r_x=\varphi_t=0$ on the reflection line $x=0$, along which we evaluate the curvature.

We compare $q=0.7$ and $q=0.8$, each with both profiles. The $q=0.7$, $f=\tanh x$ family is the baseline, with $A=0.4$ as its representative inward trajectory.

For fixed $A$, $q$, and $f$, we follow the time evolution of $r$, $m$, and $K$ on $x=0$ and examine how the curvature \(K\) depends on the shrinking areal radius \(r\). For the amplitude comparison, we fix $q$, $f$, and the initial gauge, vary $A$, and evolve each solution to $S=(t,x)=(T,0)$, with $T=5$ unless stated otherwise. The squared radius there is $\rho_S(A)=r^2(T,0;A)$, and the corresponding sphere has area $4\pi\rho_S(A)$. We also vary $T$ from $4$ to $6$ in the baseline family while retaining the same initial gauge.

\begin{figure}[tbp]
\includegraphics[width=\columnwidth]{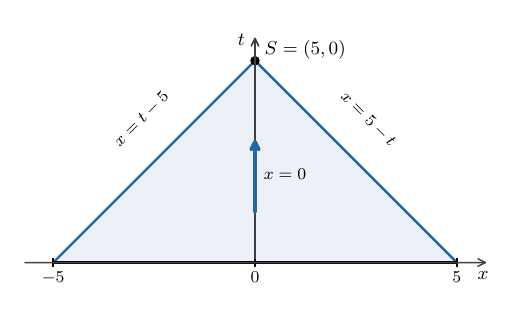}
\caption{Spacetime diagram showing the initial interval and the comparison point \(S=(t,x)=(5,0)\). The initial segment $t=0$, $-5\leq x\leq5$ determines the shaded triangle bounded by radial null curves $x=\pm(5-t)$. The vertical arrow follows $x=0$ toward increasing $t$. Solutions with different initial scalar-field amplitudes \(A\) are compared at the same coordinate point $S$.}
\label{fig:domain}
\end{figure}

Fig.~\ref{fig:domain} shows the future domain of dependence of the initial interval $[-5,5]$. The two radial null boundaries meet at \(S\). The geometry is symmetric under \(x\to -x\), and each regular point on \(x=0\) represents a sphere of nonzero area. App.~\ref{sec:initial} gives the constraints, initial data, and amplitude-response equations. App.~\ref{sec:decomposition} supplies the curvature and mass-evolution derivations, and App.~\ref{sec:numerics} describes the evolution algorithm and the reconstruction of $A_*$ and $c_*$.

\section{Curvature growth and the electric correction}
\label{sec:curvature}
\subsection{Exact curvature decomposition and lower bound}
Define the scalar-gradient invariant
\begin{equation}
 X=\nabla_a\varphi\nabla^a\varphi
  =\frac{\varphi_x^2-\varphi_t^2}{E}.
\end{equation}
It can have either sign in the general spherical solution. Einstein's equation gives $R=8\pi X$ and
\begin{equation}
 R_{ab}R^{ab}=\frac{4q^4}{r^8}-\frac{16\pi q^2X}{r^4}+64\pi^2X^2.
 \label{eq:ricci}
\end{equation}
Spherical symmetry leaves one independent Weyl amplitude $W$, with
\begin{equation}
 W=\frac{m}{r^3}-\frac{q^2}{r^4}+\frac{2\pi X}{3},\quad
 C_{abcd}C^{abcd}=48W^2.
 \label{eq:weyl}
\end{equation}
Here $C_{abcd}$ is the Weyl tensor, the trace-free part of the curvature. Using $K=C_{abcd}C^{abcd}+2R_{ab}R^{ab}-R^2/3$ and completing the square gives
\begin{equation}
 K=48W^2+\frac{320\pi^2}{3}
 \left(X-\frac{3q^2}{20\pi r^4}\right)^2+\frac{28q^4}{5r^8}.
 \label{eq:squares}
\end{equation}
Since the squared terms are nonnegative at every regular point with $r>0$, any sequence with $r\to0$ at fixed $q\ne0$ obeys
\begin{equation}
 Kr^6\geq\frac{28q^4}{5r^2}\longrightarrow\infty.
 \label{eq:no6}
\end{equation}
This conclusion does not require the time symmetry or scalar parity of our initial data. An asymptotic law $K\sim C r^{-\alpha}$ with a finite, positive constant $C$ must therefore have $\alpha\geq8$. Local exponents near six remain possible. The bound does not fix the limiting exponent to eight.

\subsection{Mass-dominated curvature scaling}
The mass contribution $m(r)^2/r^6$ contains both the explicit sixth power and the radial variation of $m$. On a contracting segment of the reflection line, the field equations give
\begin{equation}
 \frac{dm}{dr}=-2\pi r^2X,\quad
 \beta\equiv-\frac{d\ln m}{d\ln r}=\frac{2\pi r^3X}{m}.
 \label{eq:massflow}
\end{equation}
We use $m>0$, as in the evolutions considered. The parameter $\beta$ is nonnegative on $x=0$ line because $X=\varphi_x^2/E\geq0$. Substitution in the curvature decomposition yields
\begin{equation}
 K={}(48+32\beta+32\beta^2)\frac{m^2}{r^6}-(96+48\beta)\frac{mq^2}{r^7}+\frac{56q^4}{r^8}.
 \label{eq:betaK}
\end{equation}
Introduce
\begin{equation}
 \mu=\frac{mr}{q^2},\quad Q_K=\frac{K\rho^3}{48m^2}.
 \label{eq:QK}
\end{equation}
The dimensionless parameter $\mu$ controls the relative size of the mass and electric terms. The ratio $Q_K$ measures the correction to the leading mass contribution. Equation~\eqref{eq:betaK} becomes
\begin{equation}
 Q_K=1+\frac23\beta+\frac23\beta^2-\frac{2+\beta}{\mu}
    +\frac{7}{6\mu^2}.
 \label{eq:QKexact}
\end{equation}
Thus $\beta\ll1$ and $\mu\gg1$ give $K\simeq48m^2/\rho^3$. A large mass suppresses electric corrections through $\mu$, whereas small $\beta$ suppresses the scalar correction and the variation of $m^2$ with radius.
The local curvature exponent obeys the exact relation
\begin{equation}
 \alpha_{\rm eff}\equiv-\frac{d\ln K}{d\ln r}
 =6+2\beta-\frac{d\ln Q_K}{d\ln r}.
 \label{eq:alpha}
\end{equation}
For this mass-dominated mechanism, $\beta\ll1$ and $\mu\gg1$ must be accompanied by $|d\ln Q_K/d\ln r|\ll1$ to obtain a near-six exponent. A small correction $Q_K-1$ alone does not ensure a small derivative. For nearly constant $\beta$ and negligible electric terms, $m\propto r^{-\beta}$ and $K\propto r^{-(6+2\beta)}$. Mass growth therefore increases the exponent above six in this approximation. If $\beta$ decreases inward, the decrease in $Q_K$ can instead lower the exponent below six even though $m$ continues to grow.

\begin{figure}[tbp]
\includegraphics[width=\columnwidth]{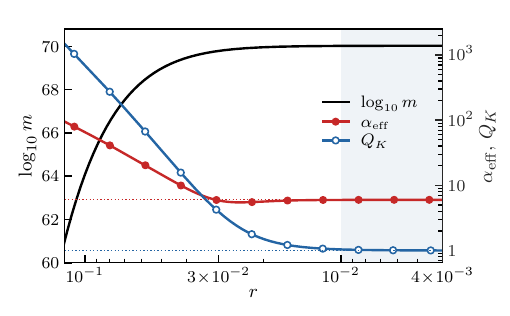}
\caption{Mass growth and curvature scaling for $A=0.4$, $q=0.7$, and $f=\tanh x$. The black solid line shows $\log_{10}m$ on the left axis. Red filled circles show $\alpha_{\rm eff}$, and blue open circles show $Q_K=Kr^6/(48m^2)$, both on the logarithmic right axis. The logarithmic radial coordinate decreases to the right. Markers identify selected data points. Red and blue dotted lines mark $\alpha_{\rm eff}=6$ and $Q_K=1$, respectively. Shading marks $0.004\leq r\leq0.01$, where the mass is nearly constant and both curvature diagnostics approach these reference values.}
\label{fig:mass}
\end{figure}

Fig.~\ref{fig:mass} shows rapid scalar-driven mass growth, consistent with mass inflation \cite{GuoJoshi}, followed by a mass plateau. Mass inflation establishes a large mass, while the subsequent plateau permits the near-six curvature exponent. Over $0.004\leq r\leq0.01$, the mass ranges from $1.05\times10^{70}$ to $1.07\times10^{70}$, while $\beta$ ranges from about $0.0025$ to $0.055$. The blue curve shows $1.0017\leq Q_K\leq1.0388$ over this interval, consistent with Eq.~\eqref{eq:QKexact} and a negligible electric contribution. On the mass plateau, the fitted curvature exponent is \(5.9955\), consistent with the predicted near-six scaling.

\subsection{Electric corrections on the mass plateau}
\label{sec:transition}

The mixed mass--charge term in Eq.~\eqref{eq:QKexact} is of relative order $\mu^{-1}$, and the pure electric term is of order $\mu^{-2}$. If the mass remains near a plateau value $m_0$, both become important when $r$ is of order $q^2/m_0$. A larger plateau mass therefore moves the electric correction to a smaller radius.

Tab.~\ref{tab:transition} gives the radii at $\mu=1$ on the numerical trajectories. For $q=0.7$ and the $\tanh$ profile, changing $A$ from $0.4$ to $0.5$ shifts this radius by more than 42 orders of magnitude. The larger plateau mass at $A=0.4$ delays the electric correction to about $5\times10^{-71}$.

\begin{table}[tbp]
\caption{Areal radius at $\mu=1$, in units of the initial RN mass parameter $M=1$. The $A=0.4$ row continues the baseline case of Fig.~\ref{fig:mass}. The four $A=0.5$ rows correspond to Fig.~\ref{fig:transition}.}
\label{tab:transition}
\begin{ruledtabular}
\begin{tabular}{cccr}
$q$ & $A$ & $f$ & $\log_{10}r\big|_{\mu=1}$\\
0.7 & 0.4 & $\tanh$ & $-70.3390$\\
0.7 & 0.5 & $\tanh$ & $-28.2092$\\
0.7 & 0.5 & $\operatorname{erf}$ & $-28.1161$\\
0.8 & 0.5 & $\tanh$ & $-22.8747$\\
0.8 & 0.5 & $\operatorname{erf}$ & $-21.0157$\\
\end{tabular}
\end{ruledtabular}
\end{table}

The curvature normalized by the electric scale is
\begin{equation}
 \mathcal K=\frac{Kr^8}{q^4}.
 \label{eq:scaled}
\end{equation}
Equation~\eqref{eq:QKexact} gives $\mathcal K=48\mu^2Q_K$. Negligible scalar corrections then give
\begin{equation}
 \mathcal K\simeq48(\mu-1)^2+8.
 \label{eq:RNshape}
\end{equation}
Since $\mu=mr/q^2$, its radial derivative satisfies $d\ln\mu/d\ln r=1-\beta$. Combining this with $\alpha_{\rm eff}=8-d\ln\mathcal K/d\ln r$ and Eq.~\eqref{eq:RNshape} gives, for $\beta\ll1$ and negligible contributions from the $\beta$-dependent terms and their radial derivatives,
\begin{equation}
 \alpha_{\rm eff}\simeq8-
 \frac{96\mu(\mu-1)}{48(\mu-1)^2+8}.
 \label{eq:RNalpha}
\end{equation}
When the mass is nearly constant and the scalar corrections and their radial derivatives are negligible, \(\mathcal K\) and \(\alpha_{\rm eff}\) take the RN expressions with the mass replaced by the plateau value \(m_0\).

\begin{figure}[tbp]
\includegraphics[width=\columnwidth]{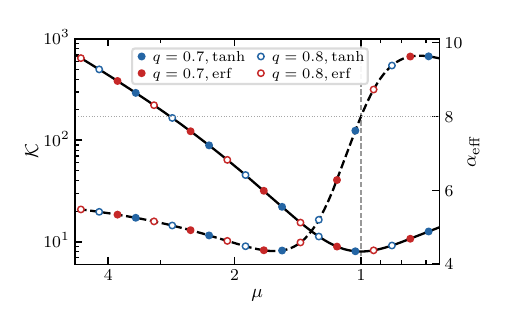}
\caption{Electric correction on the mass plateau at $A=0.5$. Blue and red circles denote the $\tanh$ and $\operatorname{erf}$ profiles, respectively. Filled and open circles denote $q=0.7$ and $q=0.8$. The black solid curve gives $\mathcal K$ (left) from Eq.~\eqref{eq:RNshape}, and the dashed curve gives $\alpha_{\rm eff}$ (right) from Eq.~\eqref{eq:RNalpha}. The vertical dashed line marks $\mu=1$. Both $r$ and $\mu$ decrease to the right.}
\label{fig:transition}
\end{figure}

At $A=0.5$, the four combinations of charge and profile reach $\mu=O(1)$ and follow the predicted curves in Fig.~\ref{fig:transition}. The mixed term first reduces the slope below six. The slope then rises through eight near the minimum of $\mathcal K$ at $\mu=1$. The identity $\alpha_{\rm eff}=8-d\ln\mathcal K/d\ln r$ relates this crossing to the local minimum of $\mathcal K$.

The exact mass definition bounds the range of validity of the constant-mass approximation. On the reflection line,
\begin{equation}
 \mu=\frac12\left(1+\frac{r^2}{q^2}+\frac{r^2r_t^2}{q^2E}\right)
 \geq\frac12\left(1+\frac{r^2}{q^2}\right).
 \label{eq:mumin}
\end{equation}
Thus the pure-charge limit $\mu\to0$ is inaccessible there. Any continuation to arbitrarily small radius must have unbounded $m$, so the constant-mass approximation must eventually fail. The crossing of eight characterizes the electric correction on this finite mass plateau. Determining the final exponent requires control of the subsequent mass growth.

\subsection{Electric transition and Planckian curvature}
\label{sec:physical_scales}
The delayed electric correction can be compared with the Planck curvature scale. Hansen, Khokhlov, and Novikov studied surfaces of Planckian Kretschmann curvature in charged black holes perturbed by a scalar field~\cite{Hansen2005}. The plateau approximation now locates the electric transition relative to that scale. Let $M_{\rm B}$ be the physical mass corresponding to the initial RN parameter $M=1$, and define $L=GM_{\rm B}/c^2$. Physical radii equal $Lr$, and the physical curvature is $K_{\rm phys}=K/L^4$. The plateau mass $m_0$ remains dimensionless, with internal mass $M_{\rm B}m_0$, and the geometric charge is $Lq$. Denoting the physical radius at $\mu=1$ by $R_{\rm e}$, and neglecting scalar corrections there, Eq.~\eqref{eq:scaled} gives
\begin{equation}
 R_{\rm e}\simeq\frac{Lq^2}{m_0},\quad
 K_{{\rm e},{\rm phys}}\simeq\frac{8m_0^8}{q^{12}L^4}.
 \label{eq:physical_transition}
\end{equation}
Using $\ell_{\rm P}=\sqrt{\hbar G/c^3}\simeq1.616\times10^{-35}\,\mathrm m$ and $K_{\rm P}=\ell_{\rm P}^{-4}$, we obtain
\begin{equation}
 \frac{K_{{\rm e},{\rm phys}}}{K_{\rm P}}
 \simeq\frac{8m_0^8}{q^{12}}\left(\frac{\ell_{\rm P}}{L}\right)^4.
 \label{eq:planck_ratio}
\end{equation}
The transition reaches Planckian curvature when
\begin{equation}
 m_0\simeq m_{0,\rm crit}
 \equiv8^{-1/8}q^{3/2}\sqrt{\frac{L}{\ell_{\rm P}}}.
 \label{eq:planck_mass_threshold}
\end{equation}
For $M_{\rm B}=10M_\odot$ and $q=0.7$, this value is about $1.37\times10^{19}$. If $m_0\gg m_{0,\rm crit}$, the electric transition lies well above Planckian curvature scale. Provided the mass plateau and small, slowly varying scalar corrections persist, the classical near-six law can then extend to Planckian curvature while electric corrections remain negligible.

The baseline trajectory illustrates the separation of these radii. With $M_{\rm B}=10M_\odot$ and $m_0\simeq1.07\times10^{70}$, the mass-dominated expression $K_{\rm phys}\simeq48m_0^2/(L^4r^6)$ reaches $K_{\rm P}$ at the physical radius
\begin{equation}
 R_{\rm P}\simeq48^{1/6}(m_0L\ell_{\rm P}^2)^{1/3}
 \simeq66\,\mathrm m.
 \label{eq:planck_radius}
\end{equation}
This corresponds to $r\simeq0.00446$, within the near-six interval in Fig.~\ref{fig:mass}, and $\mu\simeq9.7\times10^{67}$, where electric corrections are negligible. The classical electric transition instead occurs at $R_{\rm e}\simeq6.8\times10^{-67}\,\mathrm m$. At fixed $L$ and $q$, increasing $m_0$ moves this transition inward as $R_{\rm e}\propto m_0^{-1}$, while the mass-dominated Planck-curvature radius increases as $R_{\rm P}\propto m_0^{1/3}$.

For the same $10M_\odot$ normalization, all five trajectories in Tab.~\ref{tab:transition} reach $\mu=1$ above Planckian curvature. Even the smallest estimate, for $q=0.8$, $A=0.5$, and the $\operatorname{erf}$ profile, gives $K_{{\rm e},{\rm phys}}/K_{\rm P}\simeq6.3\times10^{12}$ at $R_{\rm e}\simeq1.4\times10^{-17}\,\mathrm m$. Thus a radius much larger than $\ell_{\rm P}$ need not imply sub-Planckian curvature. The ordering depends on the physical mass scale: at fixed dimensionless solution, the ratio in Eq.~\eqref{eq:planck_ratio} decreases as $M_{\rm B}^{-4}$. For this least extreme trajectory, it reaches unity at $M_{\rm B}\simeq1.6\times10^4M_\odot$.

The condition $K_{\rm phys}\ell_{\rm P}^4\sim1$ is a curvature benchmark for the classical solution. It does not determine when quantum backreaction changes that solution.

\section{Area response and the joint curvature law}
\label{sec:area}

A joint scaling law follows when $K_S\rho_S^3/(48m_S^2)\simeq1$ and $\rho_S$ responds linearly to $A$ over the same amplitude interval.

\subsection{Linear area response and the half-power law}
Define the amplitude response at fixed coordinates and fixed initial gauge by
\begin{equation}
 Z=\partial_A\rho.
 \label{eq:response}
\end{equation}

Suppose a regular branch approaches $\rho_S=0$ at $A_*$, and $-Z_S$ tends to a finite positive value $c_*$. Integration then gives
\begin{align}
 \rho_S(A)&=\int_A^{A_*}[-Z_S(a)]\,da\sim c_*(A_*-A),
 \label{eq:simplezero}\\
 r_S(A)&\sim\sqrt{c_*}\,(A_*-A)^{1/2}.
 \label{eq:half}
\end{align}
Here $\sim$ means that the ratio tends to one as $A$ approaches $A_*$ from below. In the numerical solutions, we examine whether $-Z_S$ remains positive and nearly constant while $\rho_S$ decreases as $A$ increases.

In particular, Eq.~\eqref{eq:rho} yields
\begin{equation}
 Z_{tt}-Z_{xx}=4EB\left(1-\frac{q^2}{\rho}\right)
 -\frac{2Eq^2}{\rho^2}Z,
 \label{eq:Z}
\end{equation}
where $B=\partial_A\sigma$ is the metric response. The initial radius also depends on $A$, so its radial constraint must be differentiated.

For the numerical results below, $A_*$ and $c_*$ denote reconstructed estimates of the zero and response coefficient. The reconstructed threshold differs from that of Ref.~\cite{Shao}, a difference that may depend on numerical implementation and is not pursued here.

\begin{figure}[tbp]
\includegraphics[width=\columnwidth]{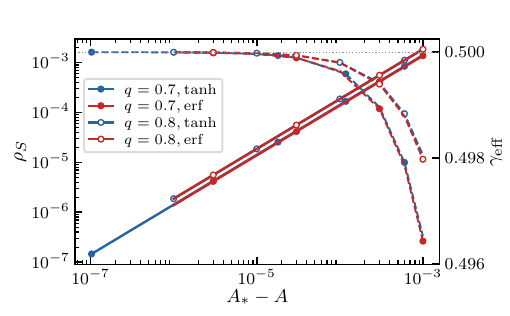}
\caption{Area response at $S=(5,0)$ for the same four charge--profile combinations as in Fig.~\ref{fig:transition}. Colors and marker fills retain the same meanings. Solid lines show $\rho_S$ (left), and dashed lines show $\gamma_{\rm eff}$ (right). Lines join the amplitude samples, with markers at selected points. The dotted line marks $1/2$. Each family uses its own $A_*$.}
\label{fig:area}
\end{figure}

The directly evaluated local radius exponent is
\begin{equation}
 \gamma_{\rm eff}=-(A_*-A)\,\partial_A\ln r_S
 =-\frac{(A_*-A) Z_S}{2\rho_S}.
 \label{eq:gamma}
\end{equation}
The linear law $\rho_S=c_*(A_*-A)$ has $\gamma_{\rm eff}=1/2$. Fig.~\ref{fig:area} compares $q=0.7$ and $q=0.8$, with the $\tanh$ and $\operatorname{erf}$ profiles at each charge.

The response coefficients vary with charge and profile, while the local radius exponents approach $1/2$ over the sampled amplitude intervals.

At the closest sampled amplitude in the baseline family, $A=0.3986179$, the full evolution gives $\rho_S\simeq1.46\times10^{-7}$ and $-Z_S\simeq1.41839$. Small area with a finite response supports the linear-area mechanism over the resolved amplitude interval.

The same test can be repeated at $S=(T,0)$ with a different observation time. For $q=0.7$ and $f=\tanh x$, the sampled times are $T=4$, $4.5$, $5$, $5.5$, and $6$, with the initial gauge held fixed. The reconstructed zero decreases from $A_*\simeq0.402433$ at $T=4$ to $0.396914$ at $T=6$, while $c_*\simeq1.42$ throughout. At each sampled time, $\gamma_{\rm eff}$ approaches $1/2$ over the resolved amplitude interval. Thus the half-power behavior persists over the sampled times even though the zero itself depends on $T$.

\subsection{Mass-normalized curvature scaling}
\label{sec:joint}
The joint scaling law requires the linear area response and mass dominance to hold at the points being compared. We therefore evaluate $\rho_S$, $m_S$, and $K_S$ together at $S=(5,0)$ for each initial amplitude. The $A=0.4$ trajectory in Fig.~\ref{fig:mass} lies outside this scan. Where $\rho_S\simeq c_*(A_*-A)$ and $Q_K(S)\simeq1$ hold together,
\begin{equation}
 \frac{K_S}{m_S^2}\simeq\frac{48}{c_*^3(A_*-A)^3}.
 \label{eq:joint}
\end{equation}

\begin{figure}[tbp]
\includegraphics[width=\columnwidth]{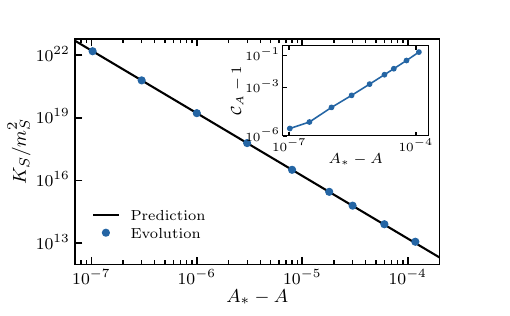}
\caption{Joint test at $S=(5,0)$ for $q=0.7$ and $f=\tanh x$. Circles show the numerical $K_S/m_S^2$, and the solid line gives Eq.~\eqref{eq:joint} with $c_*$ fixed by the reconstructed slope of $\rho_S(A)$. The inset shows the relative departure $\mathcal C_A-1$ from Eq.~\eqref{eq:joint}. Its lines join the amplitude samples.}
\label{fig:joint}
\end{figure}

A mass plateau along one trajectory does not fix the mass in solutions with different initial amplitudes. Dividing by $m_S^2$ removes this amplitude dependence from the leading curvature term. Fig.~\ref{fig:joint} compares Eq.~\eqref{eq:joint} directly with the measured $K_S/m_S^2$, using $A_*$ and $c_*$ from the area reconstruction. No coefficient is fitted to the curvature data.

Define the dimensionless ratio
\begin{equation}
 \mathcal C_A=\frac{c_*^3(A_*-A)^3K_S}{48m_S^2}
 =Q_K(S)\left(\frac{c_*(A_*-A)}{\rho_S}\right)^3.
 \label{eq:joint_ratio}
\end{equation}
Equation~\eqref{eq:joint_ratio} separates departures from linear area response from corrections to mass-dominated curvature. The joint law requires $Q_K(S)$ and $c_*(A_*-A)/\rho_S$ to be close to one over the same amplitude interval. Both exceed one in the samples, so cancellation cannot account for $\mathcal C_A$ being close to unity.

At the largest sampled $A_*-A$, the radius exponent is already close to $1/2$, but $\beta_S\simeq0.197$ leaves an appreciable scalar contribution to curvature. The inset in Fig.~\ref{fig:joint} resolves the resulting departure from the joint law: $\mathcal C_A-1$ decreases from about $0.16$ to $2.8\times10^{-6}$ over the sampled amplitude interval. Meanwhile $m_S$ falls from $7.28\times10^{72}$ to $4.34\times10^{72}$ as $A$ increases toward $A_*$, confirming the need for the mass normalization.

\section{Discussion and conclusions}
The near-six curvature law describes a mass-dominated stage after mass inflation. Equations~\eqref{eq:QKexact} and~\eqref{eq:alpha} identify its conditions: a slowly varying mass, small scalar and electric corrections, and a small radial variation of the relative curvature correction. On a plateau $m\simeq m_0$, the electric condition is $r\gg q^2/m_0$. Mass inflation establishes a large $m_0$, while the subsequent plateau permits the exponent near six. The resolved electric transition and the exact curvature bound exclude this behavior but does not determine the final exponent.

The physical-unit comparison gives a further answer to the range-of-validity question. If $m_0\gg m_{0,\rm crit}$ and the plateau and scalar-correction conditions persist, the classical near-six regime reaches Planckian curvature before electric corrections become significant. For the baseline solution with an initial RN mass scale of $10M_\odot$, this occurs at an areal radius of about $66\,\mathrm m$, while $\mu\sim10^{68}$. This conclusion concerns the classical evolution and depends on the physical mass scale assigned to the initial data.

The half-power radius law follows from a simple area zero with a finite, nonzero amplitude response. Its persistence as the observation time changes supports this mechanism beyond $S=(5,0)$, although the reconstructed zero shifts with time. Combining the linear area response with mass-dominated curvature yields $K_S/m_S^2\propto(A_*-A)^{-3}$. The numerical joint test supports this prediction for $q=0.7$ and $f=\tanh x$, with its coefficient determined from the area response.

Two extensions follow directly from these results. Searching for self-similar evolution and growing perturbations would clarify whether critical dynamics of the type studied in scalar collapse~\cite{Choptuik,Gundlach1997} accompanies the area response. Calculations of the renormalized stress tensor on RN backgrounds resolve quantum fluxes near the inner horizon~\cite{Zilberman2020}. Including the corresponding quantum stress tensor in the evolution of the scalar-field geometry would test how quantum backreaction changes the mass plateau and the ordering of the electric transition and high-curvature regime.

\begin{acknowledgments}
The author gratefully acknowledges the assistance of OpenAI's ChatGPT with numerical programming, checking the derivations, and polishing the manuscript. The author takes full responsibility for all scientific results and the content of this paper.
\end{acknowledgments}

\appendix
\section{Constraints, initial data, and amplitude response}
\label{sec:initial}
\subsection{Constraints and initial data}
\label{sec:constraints}
The Einstein constraints $H_1=H_2=0$ complement Eqs.~\eqref{eq:rho}--\eqref{eq:scalar}:
\begin{align}
 2rH_1&=\rho_{tx}-\frac{\rho_t\rho_x}{2\rho}
 +\rho_t\sigma_x+\rho_x\sigma_t+8\pi\rho\varphi_t\varphi_x,\label{eq:constraint1}\\
 \begin{split}
 2rH_2&=\rho_{xx}+\rho_t\sigma_t+\rho_x\sigma_x
 -\frac{\rho_t^2+\rho_x^2}{4\rho}\\
 &\quad+4\pi\rho(\varphi_t^2+\varphi_x^2)-E+\frac{Eq^2}{\rho}.
 \end{split}\label{eq:constraint2}
\end{align}
The first vanishes on the reflection line $x=0$. On the time-symmetric initial slice, the second determines the radius once $E(0,x)$ and $\varphi(0,x)$ are specified.

The areal radius of the initial RN background $r_{\rm RN}(x)$ is fixed implicitly by
\begin{equation}
 x^2=e^{2\kappa_+ r_{\rm RN}}
 \left(\frac{r_{\rm RN}}{r_+}-1\right)
 \left(\frac{r_{\rm RN}}{r_-}-1\right)^{\kappa_+/\kappa_-},
\end{equation}
where $r_\pm=1\pm\sqrt{1-q^2}$ and $\kappa_\pm=(r_\pm-r_\mp)/(2r_\pm^2)$. We prescribe
\begin{equation}
 E(0,x)=\frac{r_+r_-e^{-2\kappa_+r_{\rm RN}}}{\kappa_+^2r_{\rm RN}^2}
 \left(\frac{r_{\rm RN}}{r_-}-1\right)^{1-\kappa_+/\kappa_-}.
\end{equation}
With $\varphi(0,x)=Af(x)$, the time-symmetric constraint is
\begin{equation}
 r_{xx}=-\sigma_xr_x-\frac{r_x^2}{2r}-2\pi rA^2f'^2
 +\frac{E}{2r}-\frac{Eq^2}{2r^3}.
 \label{eq:initialconstraint}
\end{equation}

\subsection{Amplitude response}
\label{sec:tangent}
On the initial slice, $R_A=\partial_A r$ satisfies the derivative of Eq.~\eqref{eq:initialconstraint} at fixed gauge,
\begin{equation}
\begin{aligned}
 (R_A)_{xx}={}&\left[\frac{r_x^2-E}{2r^2}-2\pi A^2f'^2
 +\frac{3Eq^2}{2r^4}\right]R_A\\
 &-\left(\sigma_x+\frac{r_x}{r}\right)(R_A)_x-4\pi rAf'^2.
 \end{aligned}
\end{equation}
The data at $x=0$ are $R_A=(R_A)_x=0$. Set $B=\partial_A\sigma$ and $P=\partial_A\varphi$.

In addition to Eq.~\eqref{eq:Z}, differentiating Eqs.~\eqref{eq:sigma} and \eqref{eq:scalar} yields
\begin{equation}
 \begin{aligned}
 B_{tt}-B_{xx}&=-\frac{\rho_tZ_t-\rho_xZ_x}{2\rho^2}
 +\frac{\rho_t^2-\rho_x^2}{2\rho^3}Z\\
 &\quad+\frac{2E(\rho-2q^2)}{\rho^2}B
 +\frac{E(\rho-4q^2)}{\rho^3}Z\\
 &\quad+8\pi(\varphi_tP_t-\varphi_xP_x),
 \end{aligned}
\end{equation}
\begin{equation}
 \begin{aligned}
 P_{tt}-P_{xx}&=-\frac{Z_t\varphi_t+\rho_tP_t-Z_x\varphi_x-\rho_xP_x}{\rho}\\
 &\quad+\frac{\rho_t\varphi_t-\rho_x\varphi_x}{\rho^2}Z.
 \end{aligned}
\end{equation}
The initial responses are $Z=2rR_A$, $B=0$, and $P=f(x)$, with all initial time derivatives zero.

\section{Curvature decomposition and mass evolution}
\label{sec:decomposition}
\subsection{Ricci and Weyl contractions}
Write the spherical metric as $ds^2=h_{AB}dy^A dy^B+r^2d\Omega^2$, where $h_{AB}$ is the metric on the time--radial plane. Choose orthonormal directions $\hat0,\hat1$ in that plane, and write the scalar gradient as $\nabla_{\hat a}\varphi=(u,v,0,0)$. Then $X=v^2-u^2$. With $\mathcal E=q^2/r^4$, the Maxwell equation gives $|F_{\hat0\hat1}|=|q|/r^2$. Einstein's equation reads
\begin{equation}
 R_{ab}=8\pi\nabla_a\varphi\nabla_b\varphi+8\pi T^{\rm em}_{ab},
\end{equation}
and its orthonormal components are
\begin{equation}
 R_{\hat a\hat b}=\begin{pmatrix}
 8\pi u^2+\mathcal E&8\pi uv&0&0\\
 8\pi uv&8\pi v^2-\mathcal E&0&0\\
 0&0&\mathcal E&0\\
 0&0&0&\mathcal E
 \end{pmatrix}.
\end{equation}
Raising both indices changes the sign of each off-diagonal $\hat0\hat1$ entry in the contraction. Hence
\begin{equation}
\begin{aligned}
 R_{ab}R^{ab}={}&(8\pi u^2+\mathcal E)^2\\
 &+(8\pi v^2-\mathcal E)^2+2\mathcal E^2-2(8\pi uv)^2,
 \end{aligned}
\end{equation}
which gives Eq.~\eqref{eq:ricci}. The negative electromagnetic--scalar cross term is retained in the completed square.

Define $C_{\hat2\hat3\hat2\hat3}=2W$. Angular isotropy and the trace-free conditions give
\begin{alignat}{2}
 C_{\hat0\hat1\hat0\hat1}&=-2W,\quad &C_{\hat2\hat3\hat2\hat3}&=2W,\\
 C_{\hat0\hat i\hat0\hat i}&=W, &C_{\hat1\hat i\hat1\hat i}&=-W.
\end{alignat}
Here $i=2,3$, with no sum over $i$. The remaining independent Weyl components vanish by spherical symmetry and the trace-free conditions. Each nonzero component listed above contains zero or two time indices, so raising the indices leaves its contribution to the contraction positive. Summing over the Riemann index symmetries gives $C_{abcd}C^{abcd}=48W^2$. The angular sectional curvature is
\begin{equation}
 R_{\hat2\hat3\hat2\hat3}=\frac{1-(\nabla r)^2}{r^2}
 =\frac{2m}{r^3}-\mathcal E.
\end{equation}
Subtracting the angular Ricci part and adding $R/6$ yields Eq.~\eqref{eq:weyl}. Finally,
\begin{equation}
 K=48W^2+8\mathcal E^2-32\pi\mathcal E X
 +\frac{320\pi^2}{3}X^2,
\end{equation}
whose completed square is Eq.~\eqref{eq:squares}. No condition on the sign of $X$ entered the derivation.

Expanding Eq.~\eqref{eq:squares} gives
\begin{equation}
\begin{aligned}
 K={}&\frac{48m^2}{r^6}-\frac{96mq^2}{r^7}+\frac{56q^4}{r^8}\\
 &+\frac{64\pi mX}{r^3}-\frac{96\pi q^2X}{r^4}+128\pi^2X^2.
 \label{eq:expanded}
\end{aligned}
\end{equation}
For $X=0$ and $m=M$, this reduces to the exact RN curvature.

\subsection{Mass variation on the reflection line}
\label{sec:massflow}
At $x=0$, Eq.~\eqref{eq:mass} becomes
\begin{equation}
 m=\frac r2+\frac{q^2}{2r}+\frac{rr_t^2}{2E}.
\end{equation}
Differentiating in time gives
\begin{equation}
 m_t=r_t\left[\frac12-\frac{q^2}{2r^2}+\frac{r_t^2}{2E}
 +\frac rE(r_{tt}+r_t\sigma_t)\right].
\end{equation}
The radial evolution equation and the constraint~\eqref{eq:constraint2} imply
\begin{equation}
 r_{tt}+r_t\sigma_t=-\frac{r_t^2}{2r}-2\pi r\varphi_x^2
 -\frac{E}{2r}\left(1-\frac{q^2}{r^2}\right).
\end{equation}
The geometric and electric terms cancel in $m_t$, leaving $m_t=-2\pi r^2Xr_t$. Division by $r_t$ on a monotone regular segment establishes Eq.~\eqref{eq:massflow}.

\section{Numerical implementation}
\label{sec:numerics}

\subsection{Evolution and curvature evaluation}
We solve the full field equations together with their amplitude derivatives. The initial radius and its response satisfy the radial constraint and its amplitude derivative. Reflection symmetry fixes the parity of all fields, and the reported events lie within the causal domain of the initial data. The initial evolution uses spatial finite differences and Runge--Kutta time integration.

Following the inward contraction requires resolving scales far below those of the initial data. We therefore introduce local variables
\begin{equation}
 \rho=\rho_0u,\quad \sigma=\sigma_0+v,\quad
 t=t_0+\ell\tau,\quad x=\ell\xi.
\end{equation}
The scales $\rho_0$, $\sigma_0$, and $\ell$ are fixed within each patch. A centered finite-difference scheme evolves the full equations on shrinking causal patches, with fields and velocities transferred between patches. We store the accumulated scale factors as $\ln\rho_0$ and $\ln\ell$. The mass and curvature are evaluated directly through $\ln m$ and $\ln K$, using the rescaled fields and these scale factors. This avoids numerical overflow when $m$ or $K$ exceeds the floating-point range. The continuation ends when another patch no longer fits inside the available domain, not at an identified singular endpoint. For an amplitude scan at $S=(T,0)$, we evolve on the final causal cone from $t_0=T-1$ to $T$, retaining all fields, responses, and sources. The baseline comparison uses $T=5$.

For the joint test and Fig.~\ref{fig:transition}, curvature is evaluated from the metric derivatives:
\begin{equation}
\begin{aligned}
 K_{\rm geom}={}&\frac4{E^2}(\sigma_{tt}-\sigma_{xx})^2
 +\frac4{r^4}[1-(\nabla r)^2]^2\\
 &+\frac8{r^2E^2}(H_{tt}^2+H_{xx}^2-2H_{tx}^2).
\end{aligned}
\label{eq:Kgeom}
\end{equation}
Here $H_{AB}=\nabla_A\nabla_Br$ is the covariant Hessian on the time--radial plane. Centered differences supply the time derivatives and the local slope of $\ln K$ against $\ln r$. We compute $\beta$ from the scalar-field gradient using Eq.~\eqref{eq:massflow}.

\subsection{Reconstruction of the area zero and response coefficient}
\label{sec:reconstruction}
Tab.~\ref{tab:area} lists the reconstructed zero $A_*$ of $\rho_S(A)$ and its response coefficient $c_*$ used in the amplitude comparisons.

\begin{table}[tbp]
\caption{Reconstructed area zero and response coefficient at $S=(5,0)$. The listed $A_*$ values specify the subtraction $A_*-A$ used in the numerical comparisons.}
\label{tab:area}
\begin{ruledtabular}
\begin{tabular}{ccrr}
$q$ & $f$ & $A_*$ & $c_*$\\
0.7 & tanh & 0.398618003 & 1.41839\\
0.7 & erf & 0.373634048 & 1.38734\\
0.8 & tanh & 0.481011212 & 1.87195\\
0.8 & erf & 0.452950912 & 1.86507\\
\end{tabular}
\end{ruledtabular}
\end{table}
For a fixed observation time $T$, choose a regular slice $t=t_0<T$, let $\ell=T-t_0$, and define
\begin{align}
 F_{t_0}&=\rho(t_0,\ell)+\int_0^\ell\rho_t(t_0,x)\,dx,\\
 G_{t_0}&=Z(t_0,\ell)+\int_0^\ell Z_t(t_0,x)\,dx.
\end{align}
At fixed initial gauge, $G_{t_0}=\partial_A F_{t_0}$. The one-dimensional wave equation and reflection symmetry give
\begin{align}
 \rho_S&=F_{t_0}+\int_{t_0}^Tdt\int_0^{T-t}\mathcal S(t,x)\,dx,\label{eq:endpoint}\\
 Z_S&=G_{t_0}+\int_{t_0}^Tdt\int_0^{T-t}\mathcal S_A(t,x)\,dx,
 \label{eq:response_integral}
\end{align}
where $\mathcal S=-2E(1-q^2/\rho)$ and $\mathcal S_A$ is the right-hand side of Eq.~\eqref{eq:Z}. The factors follow by adding the two equal halves of the past light cone to the factor $1/2$ in the wave equation's source integral. The same symmetry reduces the initial-data contribution to $F_{t_0}$.

For each $T$, we reconstruct $A_*$ as a zero of $F_{T-1}$ and set $c_*=-G_{T-1}(A_*)$. At $T=5$, these are $F_4$ and $G_4$, evaluated on the intermediate slice $t_0=4$. The fields and responses are then continued to $S$ with all source terms retained.

The integrals of $|\mathcal S|$ and $|\mathcal S_A|$ bound the corrections to $\rho_S\simeq F_{t_0}$ and $Z_S\simeq G_{t_0}$ on the computed fields. At the closest sample in the baseline family, they are negligible compared with the resolved $\rho_S$ and $Z_S$. These integral bounds quantify the approximation of omitting the sources after $t_0$. They do not bound accumulated discretization error. The source correction requires integration over the full cone, since small $E$ on one slice does not bound $E/\rho^2$ or the metric response in Eq.~\eqref{eq:Z}.

\subsection{Departures from the scaling laws}
\label{sec:joint_checks}
\label{sec:errors}
The deviations $\gamma_{\rm eff}-1/2$ and $\mathcal C_A-1$ quantify departures from the half-power radius law and the joint curvature law, respectively. At $A=0.3986$ in the baseline family, they are approximately $-6.2\times10^{-5}$ and $6.3\times10^{-3}$. The deviation in Fig.~\ref{fig:joint} is evaluated using the reconstructed $A_*$ and does not include uncertainty in that zero.

\end{document}